\documentclass[prl,showpacs,twocolumn]{revtex4-1}
\usepackage{tabularx}
\usepackage{amsmath}
\usepackage{amssymb}
\usepackage{graphicx}
\usepackage{dcolumn}
\usepackage{bm}

\begin{document}

\title{BCS-BEC Crossover and Topological Phase Transition in 3D Spin-Orbit
Coupled Degenerate Fermi Gases}
\author{Ming Gong$^{1}$}
\author{Sumanta Tewari$^{2}$}
\author{Chuanwei Zhang$^{1}$}
\thanks{Corresponding author. Email: cwzhang@wsu.edu}

\begin{abstract}
We investigate the BCS-BEC crossover in three dimensional degenerate Fermi
gases in the presence of spin-orbit coupling (SOC) and Zeeman field. We show
that the superfluid order parameter destroyed by a large Zeeman field can be
restored by the SOC. With increasing strengths of the Zeeman field, there is
a series of topological quantum phase transitions from a non-topological
superfluid state with fully gapped fermionic spectrum to a topological
superfluid state with four topologically protected Fermi points (\textit{i.e.%
}, nodes in the quasiparticle excitation gap) and then to a second
topological superfluid state with only two topologically protected Fermi
points. The quasiparticle excitations near the Fermi points realize the
long-sought low-temperature analog of Weyl fermions of particle physics. We
show that the topological phase transitions can be probed using the
experimentally realized momentum resolved photoemission spectroscopy.
\end{abstract}
\affiliation{$^{1}$Department of Physics and Astronomy, Washington State University,
Pullman, WA, 99164 USA \\
$^{2}$Department of Physics and Astronomy, Clemson University, Clemson, SC
29634 USA}
\maketitle

Ultra-cold Fermi gases with tunable atom interaction through Feshbach
resonance \cite{Chin10} have garnered tremendous attention recently \cite%
{Review1} for their potential use as an ideal platform for emulating many
important physical phenomena. Interesting physics, including the crossover
from Bardeen-Cooper-Schrieffer (BCS) superfluid to Bose-Einstein condensate
(BEC) of molecules \cite{BCS-BEC}, universal properties in the unitary limit
\cite{Unitary}, vortices \cite{Vortices}, \textit{etc.}, have been observed
in experiments. In addition to single species Fermi atoms with equal
population in two spin states, Fermi gases with population and mass
imbalance have also been intensively studied \cite{Popimb1,Popimb2,Massimb1}%
. Here the population imbalance between two pseudo-spin states serves as an
effective Zeeman field, which have stimulated enormous experimental efforts
on searching for the Fulde-Ferrel-Larkin-Ovchinnikov state in Fermi gases
\cite{FFLO}.

The pseudospin of atoms can couple with not only the effective Zeeman field,
but also with the orbital degrees of freedom of atoms (\textit{i.e.},
spin-orbit coupling (SOC)). The recent experimental realization of SOC for
ultra-cold atoms \cite{Lin} opens a completely new avenue for investigating
the SOC physics using degenerate cold Fermi gases. In this context, it is
natural to investigate the BCS-BEC crossover physics \cite%
{BCStheory1,BCStheory2} in the presence of SOC, which is an important and
interesting problem by itself and has not been investigated previously.
Furthermore, such crossover physics is also important because the \textit{s}%
-wave superfluid, together with SOC and Zeeman field, may yield intriguing
chiral \textit{p}-wave physics \cite{Zhang08} such as Majorana fermions with
non-Abelian statistical properties \cite{Nayak}. However, such chiral
\textit{p}-wave physics may be observable only in the crossover region where
the superfluid order parameter is large and thus robust against finite
temperature effects.

In this Letter, we investigate the BCS-BEC crossover and the topological
properties of a three-dimensional (3D) uniform \textit{s}-wave superfluid in
the presence of both Zeeman field and Rashba-type of SOC. Under the mean
field approximation, we derive the superfluid gap and atom density equations
and solve them self-consistently in the BCS-BEC crossover region. Our main
results are:

(I) It is well known that the superfluid can be destroyed by the Zeeman
field beyond a critical value for a given \textit{s}-wave interaction
strength \cite{Zeeman}. We show that a finite SOC strength can restore the
superfluid pair potential back to the system even when the Zeeman field is
well above the critical value.

(II) At zero temperature, the nonzero superfluid pair potential in the
presence of SOC supports, with increasing strength of the Zeeman field, a
series of 3D topological quantum phase transitions \cite{Volovik} from a
non-topological superfluid state with fully gapped fermionic excitations to
a topological superfluid state with four protected Fermi points (\textit{i.e.%
}, gap nodes) and then to a second topological superfluid state with only
two protected Fermi points (see Fig.~3). Such 3D topological superfluids are
gapless without Majorana fermions because of the existence of Fermi points,
as opposed to the 2D fully gapped topological superfluids with Majorana
fermions \cite{Nayak}.

(III) The superfluid phases separated by the topological quantum critical
points are indistinguishable in terms of the superfluid pair potential,
which remains continuous across both transitions. On the other hand, the
superfluid phases are distinguishable in terms of the ground states which
have very different excitation spectra. The non-topological superfluid state
is a usual fully gapped \textit{s}-wave superfluid, while the topological
superfluid states are gapless even if the superfluid pair potentials are
still strictly \textit{s}-wave! These peculiar gapless topological
superfluids are new phases of matter and here we show how to identify such
phases and the corresponding quantum critical points using the
experimentally already realized momentum-resolved photoemission spectroscopy
\cite{Jin1,Jin2}.

\emph{Mean-field theory}: The system we consider is a 3D uniform \textit{s}%
-wave fermionic superfluid with the atom density $n$ and which is subject to
Rashba SOC in the $xy$ plane and a Zeeman field along the $z$ direction. The
dynamics of the Fermi gas can be described by the Hamiltonian $H=H_{0}+H_{%
\text{int}}$, where the single particle Hamiltonian $H_{0}=\sum_{\mathbf{k}%
\gamma \gamma ^{\prime }}c_{\mathbf{k}\gamma }^{\dagger }\left[ \xi _{%
\mathbf{k}}I+\alpha \left( k_{y}\sigma _{x}-k_{x}\sigma _{y}\right) +\Gamma
\sigma _{z}\right] _{\gamma \gamma ^{\prime }}c_{\mathbf{k}\gamma ^{\prime
}} $, $\gamma =\uparrow ,\downarrow $ are the pseudo-spin of the atoms, $\xi
_{\mathbf{k}}=\epsilon _{\mathbf{k}}-\mu $, $\epsilon _{\mathbf{k}}=\hbar
^{2}k^{2}/2m$ is the free particle energy, $\mu $ is the chemical potential,
$\alpha $ is the Rashba SOC strength, $I$ is the $2\times 2$ unit matrix, $%
\sigma _{i}$ is the Pauli matrix, $\Gamma $ is the strength of the Zeeman
field, and $c_{\mathbf{k}\gamma }$ is the atom annihilation operator. $H_{%
\text{int}}=g\sum_{\mathbf{k}}c_{\mathbf{k}\uparrow }^{\dag }c_{-\mathbf{k}%
\downarrow }^{\dag }c_{-\mathbf{k}\downarrow }c_{\mathbf{k}\uparrow }$ is
the $s$-wave scattering interaction with $g=4\pi \hbar ^{2}\bar{a}_{s}/m$,
and the scattering length $\bar{a}_{s}$ can be tuned by the Feshbach
resonance \cite{Chin10}.

In the mean-field approximation, the \textit{s}-wave superfluid pair
potential $\Delta =g\sum_{\mathbf{k}}\left\langle c_{\mathbf{k}\downarrow
}c_{-\mathbf{k}\uparrow }\right\rangle $ and $H_{\text{int}}=-\Delta
^{2}/g+\Delta \sum_{\mathbf{k}}(c_{\mathbf{k}\downarrow }c_{-\mathbf{k}%
\uparrow }+$ $c_{-\mathbf{k}\uparrow }^{\dagger }c_{\mathbf{k}\downarrow
}^{\dagger })$. Under the Nambu spinor basis $\Psi _{\mathbf{k}}=(c_{\mathbf{%
k}\uparrow },c_{\mathbf{k}\downarrow },c_{-\mathbf{k\downarrow }}^{\dagger
},-c_{-\mathbf{k}\uparrow }^{\dagger })^{T}$, the Hamiltonian is $H=\frac{1}{%
2}\sum\nolimits_{\mathbf{k}}\Psi _{\mathbf{k}}^{\dagger }M_{\mathbf{k}}\Psi
_{\mathbf{k}}-\Delta ^{2}/g+\sum\nolimits_{\mathbf{k}}\xi _{\mathbf{k}}$,
where%
\begin{equation}
M_{\mathbf{k}}=\left(
\begin{array}{cc}
H_{0}(\mathbf{k}) & \Delta I \\
\Delta I & -\sigma _{y}H_{0}^{\ast }(\mathbf{-k})\sigma _{y}%
\end{array}%
\right)  \label{MAT}
\end{equation}%
preserves the particle-hole symmetry. The quasiparticle excitation energy
\begin{equation}
E_{\mathbf{k}\pm }^{\lambda }=\lambda \sqrt{\xi _{\mathbf{k}}^{2}+\alpha
^{2}k_{\perp }^{2}+\Gamma ^{2}+|\Delta |^{2}\pm 2E_{0}}  \label{dispersion}
\end{equation}%
is the eigenvalue of $M_{\mathbf{k}}$, where $\lambda =\pm $ correspond to
the particle and hole branches, $E_{0}=\sqrt{\Gamma ^{2}(\xi _{\mathbf{k}%
}^{2}+|\Delta |^{2})+\alpha ^{2}k_{\perp }^{2}\xi _{\mathbf{k}}^{2}}$, and $%
k_{\perp }=\sqrt{k_{x}^{2}+k_{y}^{2}}$. For $\alpha =\Gamma =0$, Eq. (\ref%
{dispersion}) reduces to $E_{\mathbf{k}}^{\lambda }=\lambda \sqrt{\xi _{%
\mathbf{k}}^{2}+|\Delta |^{2}}$ in the standard BCS theory.

Diagonalizing the total Hamiltonian using the Bogoliubov transformation and
following the standard procedure \cite{Review1}, we obtain the gap equation
\begin{equation}
{\frac{m\Delta }{4\pi \hbar ^{2}a_{s}}}=-\Delta \sum\nolimits_{\mathbf{k}%
,\eta }\left( 1-\eta \Gamma ^{2}/E_{0}\right) f\left( E_{\mathbf{k},\eta
}^{+}\right) -{\frac{1}{2\epsilon _{\mathbf{k}}}},  \label{eq-kfas}
\end{equation}%
where $f\left( E_{\mathbf{k},\eta }^{+}\right) =\tanh (\beta E_{\mathbf{k}%
,\eta }^{+}/2)/4E_{\mathbf{k},\eta }^{+}${, }$\beta =1/k_{B}T$, $T$ is the
temperature, and $k_{B}${\ is the Boltzmann constant. Following the standard
procedure \cite{Review1}, the ultra-violet divergence at the large $\mathbf{k%
}$ in Eq. (\ref{eq-kfas}) has been regularized through subtracting the term $%
1/2\epsilon _{\mathbf{k}}$ and $a_{s}$ is defined as the renormalized
scattering length. The total number of atoms can be obtained using a similar
method \cite{Review1}
\begin{equation}
N=2\sum\nolimits_{\mathbf{k},\eta }\left[ {\frac{\eta (\alpha ^{2}k_{\perp
}^{2}+\Gamma ^{2})}{E_{0}}}-1\right] \xi _{\mathbf{k}}f\left( E_{\mathbf{k}%
,\eta }^{+}\right) +{\frac{1}{2}.}  \label{eq-n}
\end{equation}%
We self-consistently solve the gap equation (\ref{eq-kfas}) and the number
equation (\ref{eq-n}) for different parameters $\left( \alpha K_{F},\Gamma
,\nu ,T\right) $ for a fixed atom density $n$ to determine $\Delta $ and $%
\mu $. Here $\nu =1/K_{F}a_{s}$ and $K_{F}=\left( 3\pi ^{2}n\right) ^{1/3}$
is the Fermi vector for a non-interacting Fermi gas with the same density at
$\Gamma =\alpha =0$. The energy unit is chosen as the Fermi energy $%
E_{F}=\hbar ^{2}K_{F}^{2}/2m$. }

\begin{figure}[t]
\centering
\includegraphics[width=1.0\linewidth]{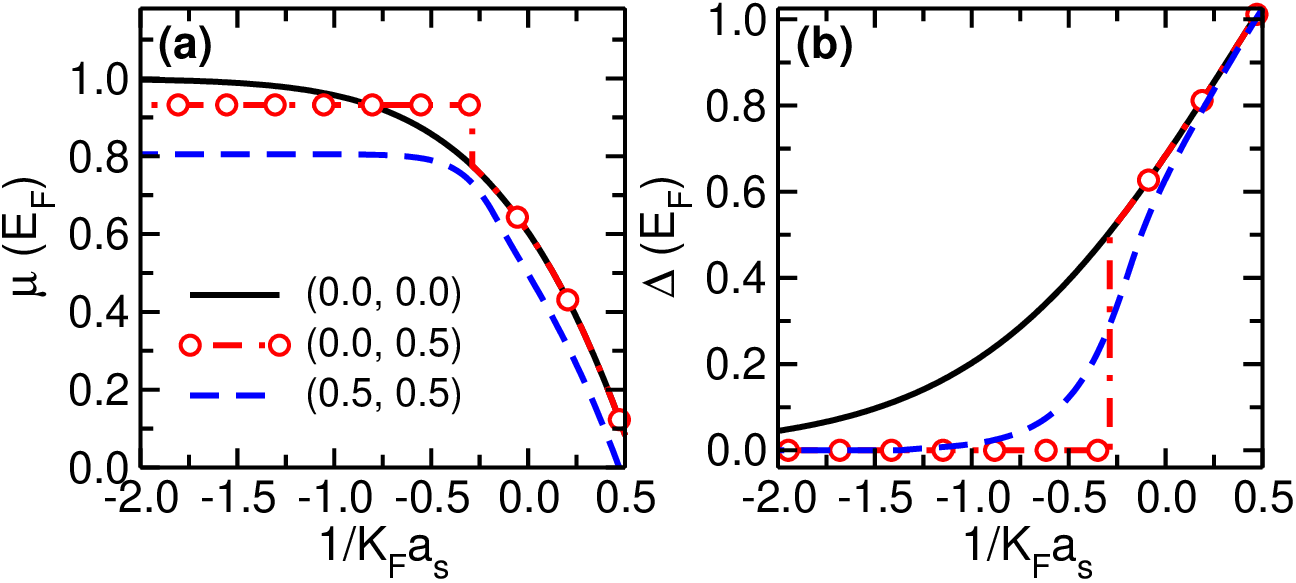}\vspace{-10pt}
\caption{(Color online) Plot of $\protect\mu $ (a) and $\Delta $ (b) versus $%
\protect\nu =1/K_{F}a_{s}$ for different parameters $(\protect\alpha %
K_{F},\Gamma )$ at $T=0$. }
\label{fig-BEC-BCS}
\end{figure}
\emph{BCS-BEC crossover}: Without SOC, it is well-known that the Zeeman
field $\Gamma $ can lift the spin degeneracy between $\mathbf{\uparrow }$
and $\mathbf{\downarrow }$, thus destroys the Cooper pairing when the Zeeman
field is larger than the pairing interaction energy \cite{Zeeman}. This
result can also be understood from Eq. (\ref{dispersion}), where the
quasiparticle excitation gap $E_{g}=\min_{\mathbf{k}}\left\vert \sqrt{|\xi _{%
\mathbf{k}}^{2}+|\Delta |^{2}}-\Gamma \right\vert =0$ for a suitable chosen $%
\mathbf{k}$ when $\alpha =0$ and $|\Gamma |>\Delta $. With SOC, the single
particle Hamiltonian $H_{0}$ has two bands and each band contains both spin $%
\mathbf{\uparrow }$ and $\mathbf{\downarrow }$ components even with a large
Zeeman field, leading to nonzero superfluid Cooper pairing between two
fermions in the same band with opposite momenta.

Such SOC induced nonzero superfluid pair potential can be clearly seen in
Fig. \ref{fig-BEC-BCS}, where we plot the change of ${\mu }${\ and }${\Delta
}${\ with respect to }$\nu =1/K_{F}a_{s}${\ for different parameters }$%
\left( \alpha K_{F},\Gamma \right) $ at\textbf{\ }$T=0$. In the BEC side,
the SOC strength $\alpha K_{F}$ and the Zeeman field $\Gamma $ do not have
significant influence on $\Delta $ because all fermion atoms form bound
molecules. Hence we mainly focus on the BCS side. At $\alpha =0$, the
superfluid pair potential is destroyed when $\Gamma >\Delta $, as expected.
In contrast, $\Delta $ and $\mu $ are independent of $\Gamma $ when $\Gamma
<\Delta $. This can be understood from the fact that, without SOC, Eq. (\ref%
{eq-kfas}) and (\ref{eq-n}) are independent of $\Gamma $ when $\Gamma
<\Delta $. Therefore there is a sudden jump of $\mu $ and $\Delta $ at $%
\Gamma =\Delta $ for $\alpha =0$, as clearly seen from Fig. {\ref%
{fig-BEC-BCS}.} We see $\Delta $ can be restored to non-zero values even for
a large\ $\Gamma $ when $\alpha K_{F}$ is switched on.

\begin{figure}[b]
\centering\includegraphics[width = 1\linewidth]{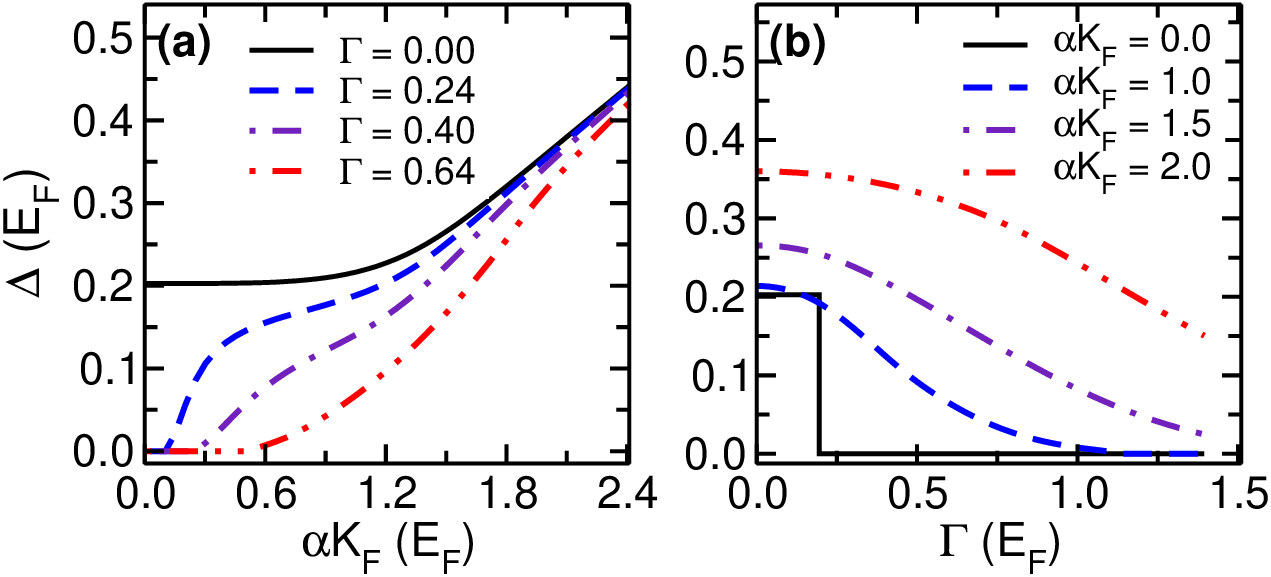}\vspace{%
-10pt}
\caption{(a) Plot of $\Delta $ with respect to $\protect\alpha K_{F}$ for
different $\Gamma $. (b) Plot of $\Delta $ with respect to $\Gamma $ for
different $\protect\alpha K_{F}$. $\protect\nu =-1$, $T=0$.}
\label{fig-alpha}
\end{figure}

In Fig. \ref{fig-alpha}a, we plot $\Delta $ with respect to $\alpha K_{F}$
on the BCS side with $\nu =-1$ and $T=0$. For other values of $\nu $ and $T$%
, the results are similar. When $\Gamma <\Delta $, the pair potential $%
\Delta $ approaches the same point (determined by the interaction strength $%
\nu $) for different $\Gamma $ when $\alpha \rightarrow 0$. $\Delta $
vanishes for $\Gamma $ beyond a critical value at $\alpha =0$, but can be
restored when $\alpha $ is nonzero. Therefore the superfluid order can still
be observed even with a large Zeeman field. In Fig. \ref{fig-alpha}b, we
plot $\Delta $ versus $\Gamma $ for different SOC. Without SOC, we observe a
sudden jump of the pair potential at $\Gamma =\Delta $, as expected. With
non-zero SOC, $\Delta $ decreases smoothly with $\Gamma $. At large $\Gamma $%
, the numerical results can be fitted with $\Delta \sim \chi \Gamma ^{-2}$
with the constant $\chi $ depending on the SOC.

\emph{Topological phase transition}: The existence of nonzero superfluid
pair potentials induced by the SOC does not automatically determine the
topological properties of the superfluid. In a 3D uniform superfluid, the
momentum is a good quantum number and the topological order of the
superfluid can be classified by the number of Fermi points \cite{Volovik} (%
\textit{i.e.}, the nodes such that $E_{\mathbf{k},-}^{+}=0$) in the
quasiparticle excitation spectrum for fixed external parameters $\left(
\alpha K_{F},\Gamma ,\nu \right) $ and at $T=0$. From Eq. (\ref{dispersion}%
), we see $E_{\mathbf{k},-}^{+}E_{\mathbf{k},+}^{+}=(\xi _{\mathbf{k}%
}^{2}-\Gamma ^{2}+\Delta ^{2}-\alpha ^{2}k_{\perp }^{2})^{2}+4\Delta
^{2}\alpha ^{2}k_{\perp }^{2}$, therefore there is always a finite gap for $%
k_{\perp }>0$ and all Fermi points must lie along $k_{z}$ axis, see Fig. \ref%
{fig-phase}a. At $k_{\perp }=0$, the quasiparticle excitation gap becomes
\begin{equation}
E_{g}=|\Gamma -\sqrt{\bar{\mu}^{2}+\Delta ^{2}}|,  \label{gap}
\end{equation}%
where $\bar{\mu}=\mu -\hbar ^{2}k_{z}^{2}/2m$.

We choose $\alpha K_{F}=E_{F}$, $\nu =-0.1$, $T=0$, and study the
topological phase transition driven by the varying $\Gamma $. In Fig. \ref%
{fig-phase}b, we plot $E_{g}=0$ in the $\left( k_{z},\Gamma \right) $ plane
and count the number of Fermi points for a fixed $\Gamma $. There are three
different phases: (A) a fully gapped non-topological superfluid state
without Fermi points when $\Gamma <\Delta $, (B) a superfluid state with
four Fermi points for each $\Gamma $ when $\Delta <\Gamma <\Gamma _{c}=\sqrt{%
\mu ^{2}+\Delta ^{2}}$ and (C) a second superfluid state with two Fermi
points for each $\Gamma $ when $\Gamma >\Gamma _{c}$. Around each Fermi
point, we have checked that the quasiparticle energy dispersion $E_{\mathbf{k%
}-}^{\lambda }$ is linear, therefore the Fermi point behaves like a Dirac
point, and is topologically protected with a topological charge $N_{a}=\pm 1$%
. Here $N_{a}$ is defined by the 3D topological invariant\textbf{\ }$N_{a}=%
\frac{1}{24\pi ^{2}}\epsilon _{\mu \nu \xi \chi }$tr$\oint\nolimits_{\Sigma
_{a}}dS^{\chi }G\frac{\partial G^{-1}}{\partial k_{\mu }}G\frac{\partial
G^{-1}}{\partial k_{\nu }}G\frac{\partial G^{-1}}{\partial k_{\xi }}$ \cite%
{Volovik}, where $G^{-1}=ik_{0}-M_{\mathbf{k}}$ is the Green function for
the quasiparticle excitation, $\Sigma _{a}$ is a 3D surface around the
isolated Fermi point $\left( 0,\mathbf{k}_{a}\right) $, $k_{0}$ is the
energy, and "tr" stands for the trace over the relevant spin indices. The
quasiparticle excitations near the Fermi points realize the long-sought
low-temperature analog of Weyl fermions of particle physics. At each of the
two critical points as a function of $\Gamma $ ($\Gamma =\Delta $ and $%
\Gamma =\sqrt{\mu ^{2}+\Delta ^{2}}$), Fermi points with opposite
topological charges merge and form a topologically trivial Fermi point with $%
N_{a}=0$, which then disappear with the Zeeman field. Note that
topologically protected Fermi points with $N_{a}=\pm 1$ in phases B and C
are protected in the sense that they cannot disappear with $\Gamma $ except
at quantum critical points (where they pair-wise merge to become
topologically trivial and disappear with the Zeeman field \cite{Volovik}).
Such a series of two topological quantum phase transitions (two quantum
critical points) is quite different from the situation encountered across a
\textit{p}-wave Feshbach resonance (\textit{i.e.}, driven by varying the
interaction strength), where a 3D gapless $p_{x}+ip_{y}$ BCS superfluid
becomes a gapped BEC at a single topological critical point at $\mu =0$ \cite%
{Victor}. In addition, the superfluid pair potential is still finite in our
system, while it vanishes in the $p_{x}+ip_{y}$ superfluid \cite{Victor}
when the quasiparticle excitation gap closes.

\begin{figure}[t]
\centering\includegraphics[width = 0.76\linewidth]{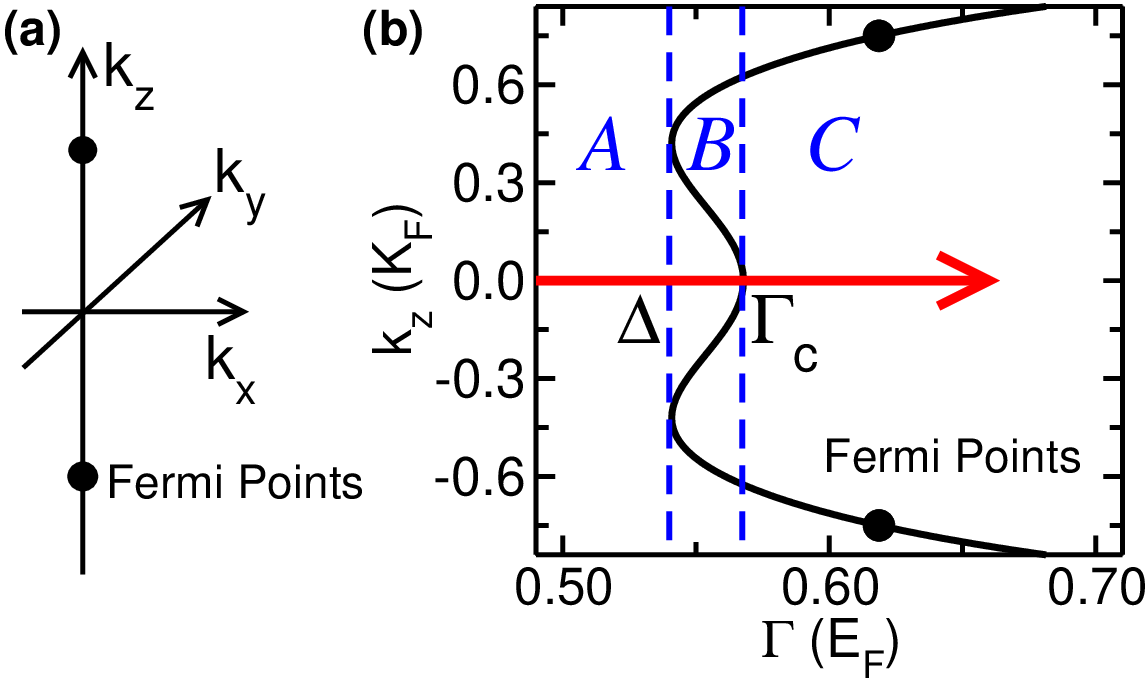}\vspace{%
-10pt}
\caption{(a) Fermi points along the $k_{z}$ axis. (b) Quantum phase
transition in a 3D uniform superfluid driven by the Zeeman field at $T=0$, $%
\protect\alpha K_{F}=E_{F}$, and $\protect\nu =-0.1$. The solid line
represents $E_{g}=0$. The change of $E_{g}$ along the red arrow is shown in
Fig. \protect\ref{fig-Eg-T}a.}
\label{fig-phase}
\end{figure}

\emph{Observation of topological phase transition}: The peculiar phases B
and C with strictly $s$-wave pair potentials, but topologically protected
Fermi points, are new phases of matter. Here we show how to identify such
phases and the corresponding topological quantum critical points by the
momentum resolved photoemission spectroscopy \cite{Jin1,Jin2}. Note that
topological quantum critical points are not easily observable in usual
experimental probes because of the absence of broken symmetries and emergent
order parameters at such transitions. In the photoemission experiments, $E_{%
\mathbf{k},-}^{+}$ for a fixed (or a small range of) $k_{z}$ can be measured
by analyzing the part of the time of flight image with the same $k_{z}$.
This measurement yields $E_{g}$ as a function of $k_{z}$ and $\Gamma $, from
which the phase diagram in Fig. \ref{fig-phase}b can be mapped out through $%
E_{g}\left( k_{z},\Gamma \right) =0$. In the following, we describe the
change of $E_{g}$ for $k_{z}=0$ and increasing $\Gamma $ (the horizontal red
arrow in Fig. \ref{fig-phase}b). The results are similar for $k_{z}\neq 0$.

For $k_{z}=0$, $E_{g}$ does not close at the boundary between phases A and
B, therefore the quasiparticle excitations are gapped non-topological when $%
\Gamma <\Gamma _{c}$ (\textit{i.e.}, A and B phases merge together). $E_{g}$
vanishes at $\Gamma _{c}$ and then reopens when $\Gamma >\Gamma _{c}$, where
the quasiparticle excitations become topologically non-trivial. In Fig. \ref%
{fig-Eg-T}a, we plot $E_{g}$ at $\mathbf{k}=0$ with respect to $\Gamma $ at $%
T=0$. The corresponding $\Delta $ and $\mu $ are also plotted. When $\Gamma $
sweeps through $\Gamma _{c}=0.56E_{F}$, $E_{g}$ first closes and then
reopens, indicating a transition from non-topological quasiparticle
excitations to gapless excitations and then to topological excitations. At $%
\Gamma _{c}$, we still observe a strong $\Delta =0.52E_{F}$, although $E_{g}$
has a node. For other fixed $k_{z}\neq 0$, the corresponding $\Gamma _{c}$
can also be determined similarly through the vanishing of $E_{g}$. Therefore
the phase diagram for the 3D topological phase transition in Fig. \ref%
{fig-phase}b can be mapped out.

\begin{figure}[t]
\centering\includegraphics[width = 1\linewidth]{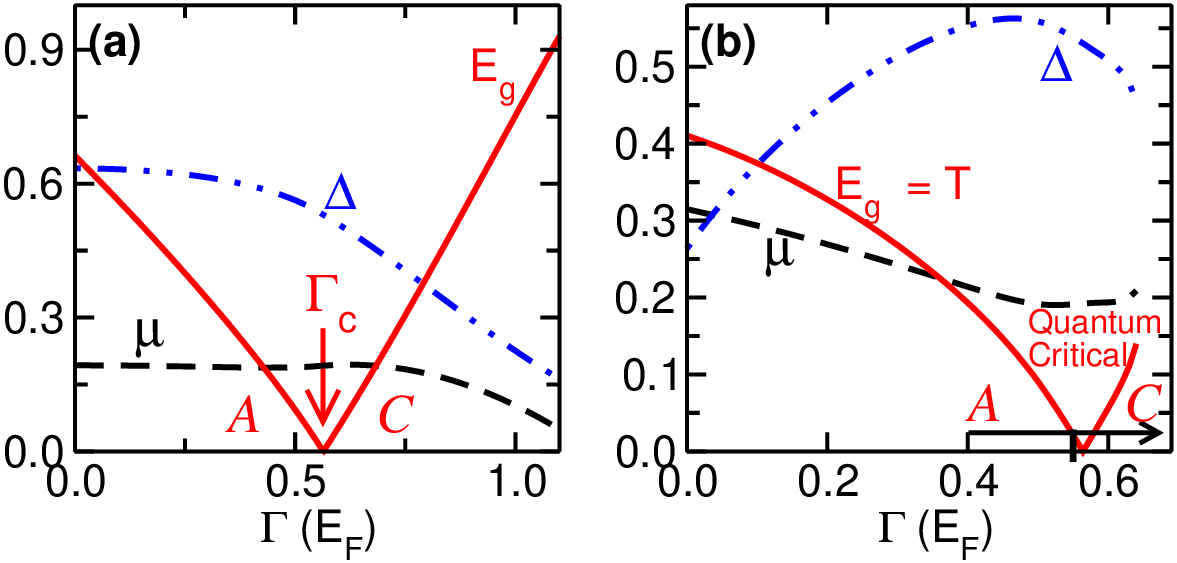}\vspace{-10pt}
\caption{(a) Plot of $E_{g}$, $\Delta $, and $\protect\mu $, with respect to
$\Gamma $ at $\mathbf{k}=0$. $\protect\nu =-0.1$, $\protect\alpha %
K_{F}=E_{F} $, and $T=0$. (b) Topological phase crossovers at finite
temperature, where the red curve is determined by Eq. \protect\ref%
{eq-boundar}. The arrow represents the phase crossovers at a fixed
temperature. $A$ and $C$ represent regions with gapped non-topological and
topological excitations, respectively. }
\label{fig-Eg-T}
\end{figure}

A realistic experiment only works at a finite temperature, which induces
thermal excitations and may destroy the superfluid. The percentage of the
thermal component in the gas depends strongly on the ratio $E_{g}/k_{B}T$.
Note that at finite temperature, there is no definite boundary between
different phases, \textit{i.e.}, there are only finite temperature phase
crossovers. Nevertheless, we still plot the phase lines in Fig. \ref%
{fig-Eg-T}b at which
\begin{equation}
E_{g}\left( \alpha ,\nu ,\Gamma ,T\right) =k_{B}T  \label{eq-boundar}
\end{equation}%
for the illustration of the topological phase crossover region. Note that $%
E_{g}$ also depends on $T$, and Eq. (\ref{eq-boundar}) should be solved
self-consistently with Eq. (\ref{eq-kfas}) and (\ref{eq-n}). Therefore the
relationship between $\Delta $ and $\Gamma $ here is quite different from
the zero temperature case in Fig. \ref{fig-Eg-T}a. In experiments, the fixed
temperature corresponds to a horizontal line (with the arrow) in the phase
diagram Fig. \ref{fig-Eg-T}b. Between phases A and C, there is a quantum
critical region where finite temperature generates many quasiparticle
excitations.

\emph{Experimental parameters}: In experiments, the Rashba SOC and Zeeman
field can be generated through the coupling between atoms and laser fields
\cite{Lin,SOC}. We consider a set of realistic parameters for $^{40}$K atom
with $n=5\times 10^{12}$ cm$^{-3}$, corresponding to the Fermi energy $%
E_{F}=h\times 3.5$ KHz. A large SOC strength $\alpha K_{F}\sim E_{F}$ is
also attainable \cite{Lin,SOC}. The effective Zeeman field $\Gamma $ from $0$
to $E_{F}$ can be obtained by tuning the intensity of laser beams \cite%
{Lin,SOC}. By tuning the \textit{s}-wave scattering length using the
Feshbach resonance \cite{Chin10}, we can vary $\Delta $ from $\sim 0.1E_{F}$
to $\sim E_{F}$. The existence of non-zero superfluid pair potentials
induced by the SOC with a large Zeeman field can be detected through the
emergence of vortices \cite{Vortices} in experiments. The topological phase
transition can be observed by detecting the change of $E_{g}$ versus $\Gamma
$ for each fixed $k_{z}$ using the momentum resolved photoemission
spectroscopy \cite{Jin1,Jin2}. From Fig. \ref{fig-Eg-T}a, we see for $%
1/K_{F}a_{s}=-0.1$ and $\Gamma =0.85E_{F}$, $E_{g}\sim E_{F}$, which is much
larger than the energy resolution in the photoemission experiment. In
experiments, the observed gap $E_{g}$ first decreases to zero and then
increases, which provides a clear signature for the transition from
non-topological to topological superfluid phases. Note that the transition
temperature $T_{c}$ from superfluid to normal states in the presence of SOC
should be larger than that for the superfluid without SOC \cite{BCS-BEC}
because of the enhanced superfluid order parameter $\Delta $ (see Fig. \ref%
{fig-alpha}a) \cite{Zhai}. The temperature needed for the observation of the
topological phase crossover in Fig. 4b is clearly lower than $T_{c}$, which
is nonzero on both sides of the critical $\Gamma _{c}$ because of the
existence of the non-vanishing $\Delta $.

\emph{Summary}: In summary, we study the BCS-BEC crossover and the
topological phase transitions in 3D uniform spin-orbit coupled degenerate
Fermi gases. The predicted SOC induced \textit{s}-wave superfluid opens new
possibilities for generating and observing many new topological phenomena in
Fermi gases. The observation of the 3D topological phase transitions using
the experimentally already realized photoemission spectroscopy provides a
critical first step for searching for non-trivial topological superfluid
states (which in 2D support Majorana fermions and the associated non-Abelian
statistics \cite{Nayak,Sau}) in cold atom \textit{s}-wave superfluids, which
are of not only fundamental but also technological importance.

\emph{Note:} After we submitted the paper, two manuscripts appeared in arXiv
\cite{Zhai,Hui}, where the BCS-BEC crossover with SOC (but without Zeeman
fields) is discussed.

\emph{Acknowledgement} This work is supported by DARPA-YFA
(N66001-10-1-4025), ARO (W911NF-09-1-0248), and DARPA-MTO (FA9550-10-1-0497).

\end{document}